\documentclass[ preprint, a4paper,pra,onecolumn,showpacs]{revtex4}
\usepackage{amsfonts}
\usepackage{amsmath}
\usepackage{amssymb}
\usepackage{graphicx}

\input{tcilatex}

\begin{document}

\title{Achievement of Quantum Degeneracy in a Na-QUIC\ trap in Brazil: an \ 
\textit{in situ} observation.}
\author{K. M. F. Magalh\~{a}es, S. R. Muniz, E. A. L. Henn, R. R. Silva, L.
G. Marcassa and V. S. Bagnato}
\affiliation{Instituto de F\'{\i}sica de S\~{a}o Carlos \\
Universidade de S\~{a}o Paulo \\
Cx. Postal 369 - CEP 13560-370, Brazil}
\received{23 November 2004}

\begin{abstract}
Using a system composed of a Quadrupole and Ioffe Configuration (QUIC) trap
loaded from a slowed atomic beam, we have performed experiments to observe
the Bose-Einstein Condensation of Na atoms. In order to obtain the atomic
distribution in the trap, we use an \textit{in situ} out of resonance
absorption image through a probe beam, to determine temperature and density.
The phase space density ($D$) is calculated using the density profile and
the temperature. We have followed $D$ as a function of the final evaporation
frequency. The results show that at 1.65 MHz we crossed the value for $D$
expected to correspond to the critical point to start de Bose-Condensation
of the sample. Due to the low number of atoms remaining in the trap at the
critical point, the interaction produces minor effects and therefore an
ideal gas model explains well the observations. We analyze the obtained low
number in terms of efficiency of evaporation. The utility of an \textit{in
situ} detection is illustrated by measuring the harmonic gas pressure of the
trapped gas in the route to condensation.
\end{abstract}

\keywords{Bose-Einstein Condensation; \textit{In situ} observation; Sodium
atoms; Thermodynamics}
\pacs{03.75.Hh; 32.80.Pj}
\maketitle

\section{Introduction}

Bose-Einstein Condensation (BEC) is a field of research with vast interest
mainly because it touches several subjects in physics. As a phase transition
from a classical gas to a coherent matter-wave sample, it provides a large
variety of interest in thermodynamic. As a heavy occupation of phase space,
BEC is a nice object for quantum statistic. As a spontaneous symmetry
breaking, it shows interest in quantum field theory. Many phenomena related
to BEC in the past, like superfluidity and superconductivity can now be
investigated in a controlled way. The field of atomic quantum optics has
also presented many interesting possibilities related to BEC. Finally, BEC
has opened up new windows into the quantum world in general.

One of the interesting possibilities to be investigated using a BEC is the
creation of condensate samples in a non ground state configuration. Such a
non-equilibrium system can be formed if one first obtains the ground state
condensate followed by a resonant pumping transfer to an excited state. Such
schemes have been theoretically investigated by our group \cite{yukalov} and
shall be one of the goals of our experiments in the future. Another
interesting possibility is the investigation of BEC from the thermodynamic
point of view considering thermodynamic variables \cite{vitor}. A sample of
such investigation is presented in this paper.

In this paper we report the first realization of BEC in Na atoms in our
laboratory. We have followed, using absorption images, the temperature and
density of atoms within the trap up to our optical resolution. Using these
parameters we have followed the phase-space path that takes the sample to
quantum degeneracy. We start presenting our experimental setup followed by
the results and discussion. An application of our \textit{in situ}
observation is the determination of the thermodynamic parameters and
evolution as the atoms are cooled to degeneracy, which is presented in the
last section.\bigskip 

\FRAME{ftbpFU}{5.2935in}{2.9897in}{0pt}{\Qcb{Schematic diagram of the
experimental system. Atoms emerging from an effusive oven are decelerated
and trapped in a MOT aligned with slowing tube. The magnetic field profile
is presented in the figure detail, where z in given in centimeters and B in
tesla.}}{}{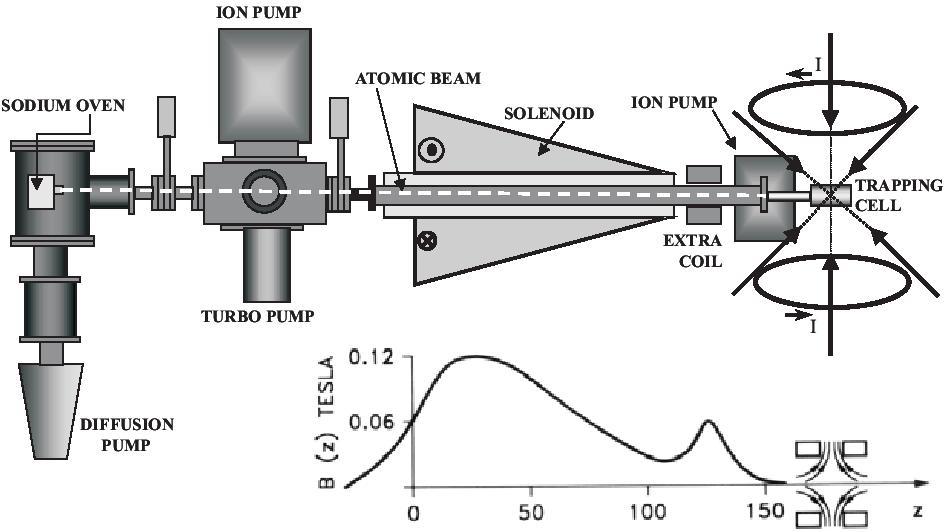}{\special{language "Scientific Word";type
"GRAPHIC";maintain-aspect-ratio TRUE;display "USEDEF";valid_file "F";width
5.2935in;height 2.9897in;depth 0pt;original-width 9.8545in;original-height
5.5417in;cropleft "0";croptop "1";cropright "1";cropbottom "0";filename
'figure1.jpg';file-properties "XNPEU";}}

\section{Experimental setup}

In order to generate the conditions to attain BEC in Na, it is necessary to
start with a thermal beam. We have loaded a Magnetic Optical Trap (MOT) from
a slow atomic beam, followed by the necessary steps to BEC \cite{becs}. Our
experimental system is presented in figure 1. An effusive sodium beam is
decelerated in a tapered solenoid employing the dark-SPOT Zeeman tuning
technique \cite{Miranda-PRA}. The process ends with a continuous flux of
slow atoms emerging from the solenoid in the lower hyperfine state. Although
the atoms are decelerated in the cycling transition $3S_{1/2}(F=2)%
\longrightarrow 3P_{3/2}(F^{\prime }=3)$, as they approach the end of the
slowing solenoid, the lower amplitude of the field as well as the
configuration of the field lines create adequate conditions to optically
pump the atoms to the $3S_{1/2}(F=1)$ hyperfine ground state. At this point
the deceleration process stops abruptly and the slowed atoms migrate out of
the solenoid without interacting with the slowing laser, travelling about 40
cm before being captured by the beams of a MOT. 

Three ring dye lasers (Coherent-699) provide the laser beams for the
slowing, trapping and repumping transitions. All lasers are frequency
stabilized and externally locked to the appropriated atomic transition,
using vapor cell and saturated absorption signal. The slower laser is tuned
to about 50 MHz to the red of the $3S_{1/2}(F=2)\longrightarrow
3P_{3/2}(F^{\prime }=3)$. The two other lasers are used to produce the
trapping and repumping frequencies of the MOT. The trapping frequency  is
tuned -5 MHz from $3S_{1/2}(F=2)\longrightarrow 3P_{3/2}(F^{\prime }=3)$
transition while the repumping light is tuned to $3S_{1/2}(F=1)%
\longrightarrow 3P_{3/2}(F^{\prime }=2)$ transition.

The MOT operates as a \textit{dark-SPOT} \cite{KetterlePRL}, where the
trapped atoms are confined in a hyperfine ground state which does not
interact with the trapping laser frequencies. To create the \textit{dark-SPOT%
} we placed a physical obstacle to block the central region of the repumper
laser beam. In such scheme, the atoms located in the dark region are rapidly
pumped to the lower energy ground state where they do not interact with the
lasers, allowing to produce a dense sample of ground state atoms. This is a
very important point to make an efficient transfer of atoms from the MOT\
phase to the magnetic trap.\bigskip 

\FRAME{ftbpFU}{3.4022in}{2.9888in}{0pt}{\Qcb{Configuration of the QUIC coils
of our experimental system. The coils for quadrupole field and the third
coil for Ioffe is in conical shape to improve the optical access. }}{}{%
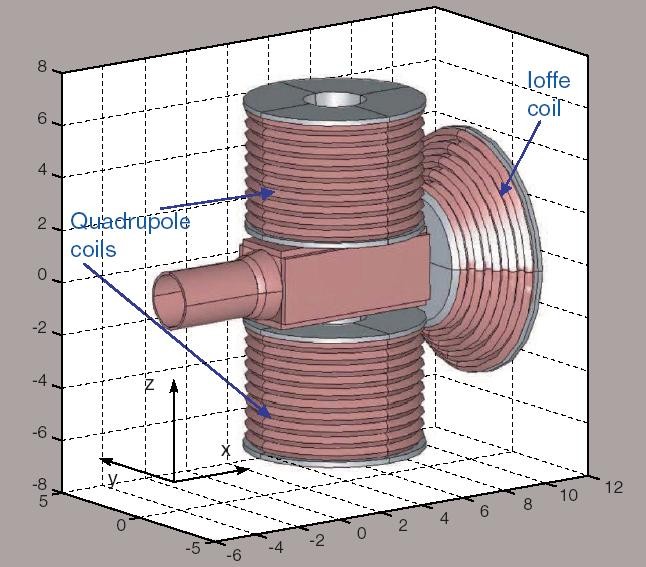}{\special{language "Scientific Word";type
"GRAPHIC";maintain-aspect-ratio TRUE;display "USEDEF";valid_file "F";width
3.4022in;height 2.9888in;depth 0pt;original-width 6.7291in;original-height
5.9067in;cropleft "0";croptop "1";cropright "1";cropbottom "0";filename
'figure2.jpg';file-properties "XNPEU";}}\bigskip 

The magnetic trap is the Quadrupole and Ioffe Configuration (QUIC)\cite{quic}%
, whose coils arrangement is presented in figure 2. The system is composed
of the two main coils (quadrupole coils) and an Ioffe coil. Each of the
coils were winded using copper tube (with a diameter of 1/8 inch) and the
number of turns in each coils is 50 for quadrupole and 30, for the Ioffe
coil. For a quadrupole current of 230 A and the Ioffe of 220 A, the produced
trap field in the Ioffe axis as well as the field contour lines in the plane
x-z are shown in figure 3a and b, respectively. The QUIC trap confining
potential is close to an harmonic oscillator, in our case, with the
following frequencies: $\omega _{\text{x}}=2\pi $ x 36.7 $Hz$; $\omega _{%
\text{y}}=2\pi $ x 120.8 $Hz$ and $\omega _{\text{z}}=2\pi $ x 159.6 $Hz.$

\FRAME{ftbpFU}{1.9043in}{2.9897in}{0pt}{\Qcb{The produced trap field for a
quadrupole current of 230 A and the Ioffe of 220 A. (a) Field profile in the
Ioffe axis and (b) field contour lines on the plane x-z, in which each value
correspond to the field in Gauss.}}{}{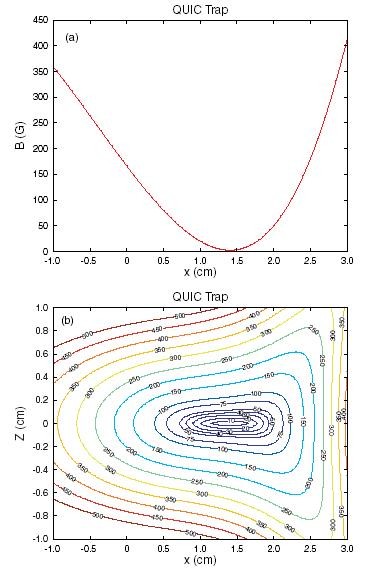}{\special{language
"Scientific Word";type "GRAPHIC";maintain-aspect-ratio TRUE;display
"USEDEF";valid_file "F";width 1.9043in;height 2.9897in;depth
0pt;original-width 3.8017in;original-height 6.0001in;cropleft "0";croptop
"1";cropright "1";cropbottom "0";filename 'figure3.jpg';file-properties
"XNPEU";}}\bigskip 

Once the atoms are in magnetic trap the evaporation process starts \cite%
{evaporation}. Through a one photon RF transition $3S_{1/2}(F=1,m_{F}=-1)%
\longrightarrow 3S_{1/2}(F=1,m_{F}=0)$, the hottest atoms are transferred to
the untrapped state carrying away energy. The RF determines the magnetic
field at which position atoms undergo a spin-flip transition, been ejected
from the trap. Thus the remaining atoms rethermalize through elastic
collisions to lower temperatures. Therefore, sweeping the RF from a top
value down, we force the evaporation until reaching the critical temperature
to observe BEC. During the evaporation stage, the atoms must be shielded
from any resonant light, that may cause optical transition destroying the
sample. The RF antenna is a coil placed close to the cell having 5 turns of
wire in a rectangular frame of $2.5$ x $7$ cm. In our experiment we use two
stages of evaporation: first on the quadrupole trap after compression and
the final evaporation on harmonic profile of the QUIC when Ioffe coil has
full current.

Our transference procedure consists of loading the MOT with about 10$^{9}$
atoms and initially transfer then to the quadrupole trap followed by
compression during 2 seconds. After final compression in quadrupole trap, at
a field gradient of 230 Gauss/cm, RF is applied to promote evaporation
during 5 seconds from 100 to 30 MHz. Then after this first stage of
evaporation, the Ioffe coil is turned on gradually during 2 seconds,
producing the final field configuration where the evaporation procedure is
completed with RF from 30 MHz to final frequency (1.65 MHz to reach BEC). 

At different stages of the cooling process, either a time of flight (TOF)
measurement can be performed to obtain the temperature or the \textit{in
situ }absorption imaging can be performed similarly to experiments already
done \cite{hulet}. Our imaging system, composed of lenses and a spatial
filtered laser beam, allows us to obtain the absorption image with a
resolution of 5 $\mu $m. We have obtained absorption image with the magnetic
field on. The image is normally obtained together with interferences and
remaining defects of the optics. Those imperfections can be cleaned out
using a Fourier Transform procedure to remove high frequency structures on
the image. The full time sequence of the magnetic trap loading and
evaporation as well as imaging timing is presented in figure 4.

\bigskip \FRAME{ftbpFU}{3.7516in}{2.9888in}{0pt}{\Qcb{The full time sequence
of the magnetic trap loading and imaging.}}{}{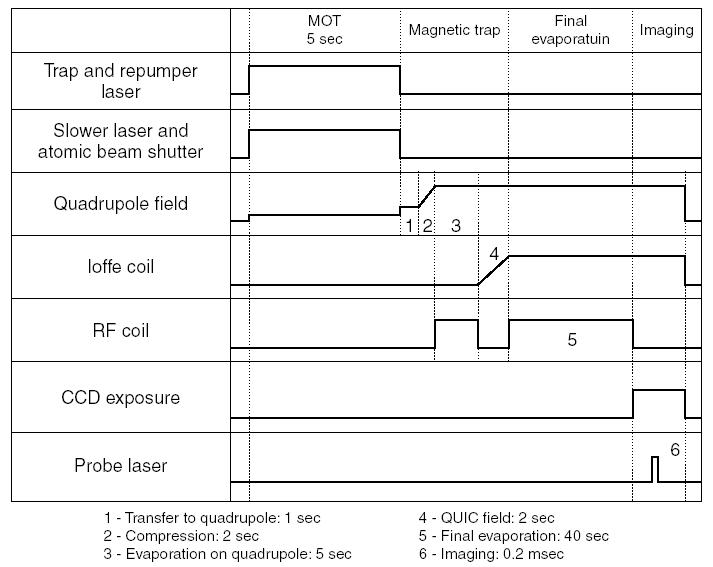}{\special{language
"Scientific Word";type "GRAPHIC";maintain-aspect-ratio TRUE;display
"USEDEF";valid_file "F";width 3.7516in;height 2.9888in;depth
0pt;original-width 7.427in;original-height 5.9067in;cropleft "0";croptop
"1";cropright "1";cropbottom "0";filename 'figure4.jpg';file-properties
"XNPEU";}}

\section{Results and discussion}

Before starting the presentation of our results we would like to present the
overall sequence of absorption images given to the reader the general idea
of the experiment. On figure 5 on the right hand side we show the typical
size of the atomic cloud and the number of atoms. On the left hand side we
have the typical temperature for each stage. In brief, the sample starts
with 10$^{9}$ atoms at 500 $\mu K$ and ends at 70 nK with a few thousand
atoms in the condensate.

\FRAME{ftbpFU}{2.2788in}{2.9888in}{0pt}{\Qcb{Sequence of absorption images
given the general idea of our experiment. On the right hand side the typical
size of the atomic cloud and the number of atoms. On the left hand side we
have the typical temperature for each stage.}}{}{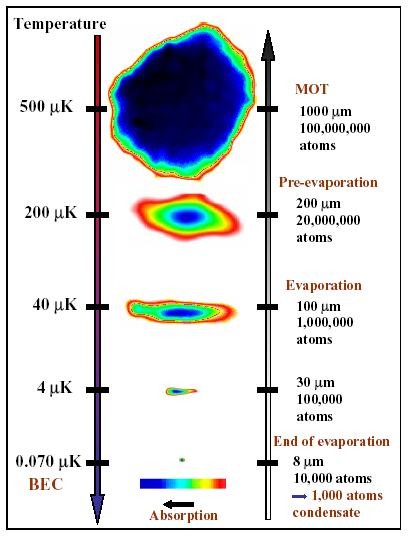}{\special%
{language "Scientific Word";type "GRAPHIC";maintain-aspect-ratio
TRUE;display "USEDEF";valid_file "F";width 2.2788in;height 2.9888in;depth
0pt;original-width 4.2601in;original-height 5.604in;cropleft "0";croptop
"1";cropright "1";cropbottom "0";filename 'figure5.jpg';file-properties
"XNPEU";}}

When we start, the atoms are very much spreaded around the quadrupole trap.
Starting the evaporation the Na atoms concentrate towards the center of the
trap and the asymmetric geometry of the magnetic field shows up on the
atomic cloud. Just after reaching BEC, the limit of resolution of our
optical system appears to start to become important.

These results were obtained using the relation between the images and the
magnetic trap geometry. The QUIC potential can be expresses as

\begin{equation}
U(\text{x,y,z})=\frac{1}{2}m(\omega _{\text{x}}^{2}\text{x}^{2}+\omega _{%
\text{y}}^{2}\text{y}^{2}+\omega _{\text{z}}^{2}\text{z}^{2})
\end{equation}

and the atomic density is found to be in good approximation with

\begin{equation}
n\left( \text{x,y,z}\right) =A\left( \text{N},\omega ,\text{T}\right) \exp
\left( -\frac{1}{2}(\frac{\text{x}^{2}}{\sigma _{\text{x}}^{2}}+\frac{\text{y%
}^{2}}{\sigma _{\text{y}}^{2}}+\frac{\text{z}^{2}}{\sigma _{\text{z}}^{2}}%
)\right) 
\end{equation}

Through absorption imaging one can obtain the values of $\sigma _{\text{x}}$
and $\sigma _{\text{z}}$ as well as the peak density at the trap center $%
n_{0}$, using a simple integration. Considering $\sigma _{\text{x}}$ and $%
\sigma _{\text{z}}$ one obtain the temperature of the cloud, since a hotter
gas spread more in space according to kT=m$\omega _{\text{z}}^{2}\sigma _{%
\text{z}}^{2}$. Once we obtain $n_{0}$ and T, the peak phase space density $D
$ is evaluated. Considering $D=n_{0}\Lambda ^{3}$ where $\Lambda =\frac{h}{%
\sqrt{2\pi mk\text{T}}}$ is the thermal de Broglie wavelength, as the peak
phase space density, one obtain for Na that

\begin{equation}
D=\frac{4,84\text{x}10^{-23}n_{0}}{\text{T}^{\frac{3}{2}}}
\end{equation}

for $n_{0}$ in cm$^{-3}$ and T in Kelvin.

According to previous work \cite{vsb1987}, once $D$ reaches the value 2.612,
the condensation starts to take place. For each final evaporation frequency
we obtain $D$ and the result is presented in figure 6 where the peak phase
space density evolution can be followed and for the final RF frequency of
1.65 MHz, the Bose-Einstein Condensation is achieved. At this point we
obtain $D\sim 14$, but still most part of the atoms are in the thermal
cloud. Since at this point we have reached our resolution limit, thus we
just overestimate the temperatures and underestimate the peak density, thus $%
D$ must be larger than the reported value. We should observe that only a few
thousand atoms are predicted to be in the condensate.

\FRAME{ftbpFU}{4.0517in}{2.9905in}{0pt}{\Qcb{The peak phase space density as
a function of final RF. \ For the final RF at 1.65 MHz, the Bose-Einstein
Condensation is achieved.}}{}{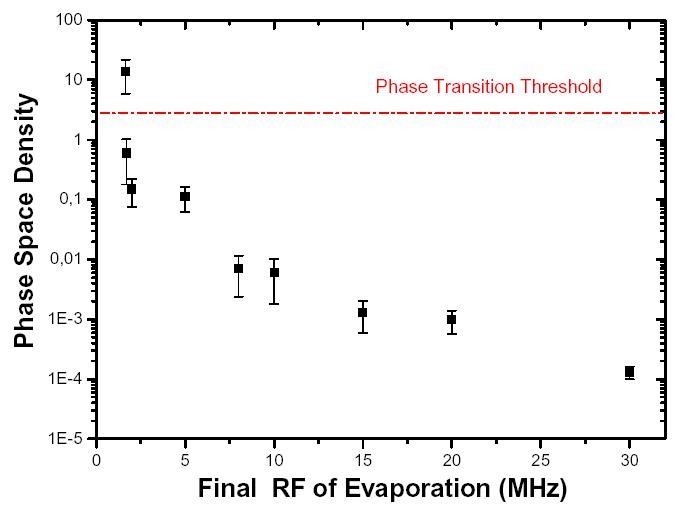}{\special{language "Scientific
Word";type "GRAPHIC";maintain-aspect-ratio TRUE;display "USEDEF";valid_file
"F";width 4.0517in;height 2.9905in;depth 0pt;original-width
7.229in;original-height 5.3229in;cropleft "0";croptop "1";cropright
"1";cropbottom "0";filename 'figure6.jpg';file-properties "XNPEU";}}\bigskip 

According to characteristic values of our trap, the ground state size should
have $\sigma _{\text{x}}^{0}$ $\sim $ $2.4$ $\mu $m, $\sigma _{\text{y}}^{0}$
$\sim $ $1.32$ $\mu $m and $\sigma _{\text{z}}^{0}$ $\sim $ $1.14$ $\mu $m,
if interactions are not considered. Using those values and the measured
number of atoms in crossing the BEC line, we obtain the critical temperature
Tc $\sim $ 61 nK which is very close to the measured 70 nK$.$

This kind of analysis is only valid if the system is close to an ideal gas.
To verify if that is true we must compare the atomic energy of the ground
state with the interaction energy at the critical point. The ground state is 
$\varepsilon _{0}$ $\sim $ $\frac{3}{2}\hbar \omega _{0}$ ($\omega _{0}=\sqrt%
[3]{\omega _{\text{x}}\omega _{\text{y}}\omega _{\text{z}}}$), which
correspond to 6.7 nK. On the other hand in the Thomas-Fermi approximation 
\cite{Thomas-Fermi}, the interaction energy/particle is given by $%
Energy/particle$ $\sim \frac{2}{7}n_{0}\tilde{U}$, where $\tilde{U}=\frac{%
4\pi \hbar ^{2}a}{m}$ with $a$ = 52(5) $a_{0}$ \cite{rev mod phys}, which is
the scattering length. In the present case, $Energy/particle$ $\sim $2 nK
and therefore we can neglect the interactions and consider the system as an
ideal gas. 

The small number of atoms attained in our BEC can be justified through of an
investigation of the variation of the number of atoms as a function of
temperature. The obtained result is presented in the figure 7. As the
evaporation on QUIC trap progress, two distinct regimes are observed. In a
first regime N$\propto $T$^{1.37}$, while for the second N$\propto $T$^{0.4}$%
. Such dependences can explain the low achieved number. Efficient
evaporation requires an elastic collision rate to be larger than the loss
rate. For continued cooling, the so called runaway evaporation, the elastic
collision rate must be maintained despite the atom number reduction. The
dependence of the atom number with temperature can be written as N$\propto $T%
$^{s}$. If s\TEXTsymbol{>}1, we have not reached the runaway evaporation
condition and we lose more atoms than necessary. If s\TEXTsymbol{<}1 the
condition for runaway evaporation is fulfilled, and better evaporative
efficiency is obtained.

\FRAME{ftbpFU}{4.0395in}{2.9905in}{0pt}{\Qcb{As the evaporation progress,
two distinct regimes are observed. In a first regime N$\propto $T$^{1.37}$,
while for the second N$\propto $T$^{0.4}$}}{}{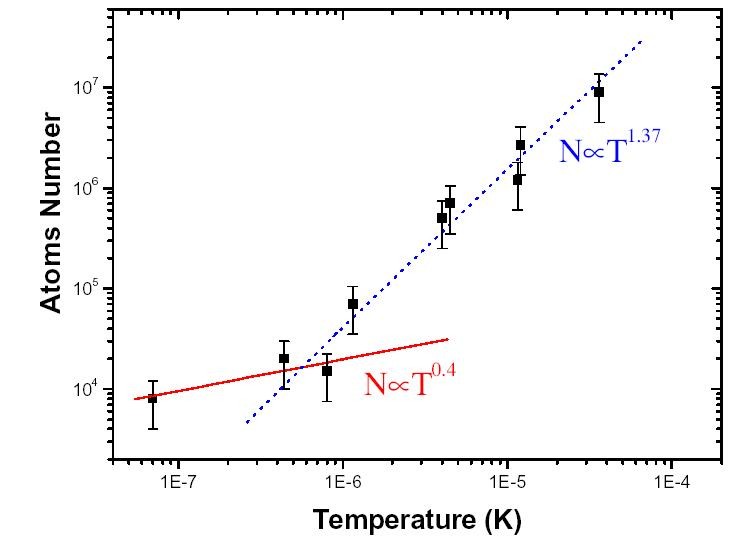}{\special{language
"Scientific Word";type "GRAPHIC";maintain-aspect-ratio TRUE;display
"USEDEF";valid_file "F";width 4.0395in;height 2.9905in;depth
0pt;original-width 7.6873in;original-height 5.6775in;cropleft "0";croptop
"1";cropright "1";cropbottom "0";filename 'figure7.jpg';file-properties
"XNPEU";}}\bigskip 

In our case we observe that runaway evaporation is only obtained at the
final stage of the evaporation process. Because of this, most part of the
atoms are expelled from the trap remaining only the coldest part and this
costs a large number of atoms, which explains the BEC with such small
number. We have tried to optimize the first part of evaporation to improve
the final result, however we have not yet been able to accomplish such
improvement. Most probable reason is that for obtaining such improvement we
have to slow down the RF sweeping speed, but in this case the lifetime of
the trapped atoms due to background collision become a limiting fact on the
process. Some modifications on our system must to be done to solve this
problem.

\section{Evaluating the harmonic pressure during the cooling down process}

Little attention has been paid to the thermodynamic of the cold trapped gas
system. For this system, usual pressure and volume are no longer
thermodynamic variables, since the confining potential is not homogeneous.
Different from the rigid walls container, atoms trapped in a magnetic field
interacts (with different amplitudes) everywhere in the potential. This
results in a non homogeneous spatial distribution of gas. In this sense we
can not carry on a thermodynamic analysis using conventional pressure and
volume to investigate the confined gas thermodynamics. It is necessary to
redefine the intensive and extensive variables. In a recent paper, V.
Romero-Rochin \cite{vitor} has investigated a confined gas in a three
dimensional harmonic potential and shown that pressure and volume are well
replaced by new variables. The extensive variable "volume " $V$\ is now
replaced by the inverse cube of the harmonic frequency $(V=\omega _{0}^{-3})$%
. There is also a conjugated intensive variable, which plays the same role
of usual pressure and is responsible for the mechanical equilibrium of the
gas. Both variables work as \ "volume" and \ \ "pressure" in the
thermodynamical sense. Considering a system composed of N particle in
equilibrium at temperature T, the thermodynamic treatment reveals that the
state equation for such system, i. e., the combination of redefined $V$ and $%
P$ gives\cite{vitor}:

\begin{equation}
P=\frac{2}{3}\omega _{0}^{3}\langle U_{ext}\rangle 
\end{equation}%
where $\langle U_{ext}\rangle $ represents an average integration over the
external potential. This pressure is referred as harmonic pressure to
distinguish the conventional pressure for gas in a box. In an explicit form,
if the gas is subjected to the potential as Eq. (1) has a density profile $%
n\left( \text{x,y,z}\right) $ as in Eq. (2) , one gets

\begin{equation}
P=\frac{2}{3}\omega _{0}^{3}\int d\text{x}d\text{y}d\text{z }n\left( \text{%
x,y,z}\right) \frac{1}{2}m(\omega _{x}^{2}\text{x}^{2}+\omega _{y}^{2}\text{y%
}^{2}+\omega _{z}^{2}\text{z}^{2})
\end{equation}

In our experiment we have followed \textit{in situ} $n\left( \text{x,y,z}%
\right) $ and knowing well the potential parameters we have followed $P$ as
a function of T. Since $\int d$x$d$y$d$z$n\left( \text{x,y,z}\right) =$N one
find that for the Gaussian density distribution, one obtain

\begin{equation}
\frac{P}{\text{N}C}=(\omega _{\text{x}}^{2}\sigma _{\text{x}}^{2}+\omega _{%
\text{y}}^{2}\sigma _{\text{y}}^{2}+\omega _{\text{z}}^{2}\sigma _{\text{z}%
}^{2})
\end{equation}%
where $\sigma _{\text{x}}$, $\sigma _{\text{y}}$ and $\sigma _{\text{z}}$
depend on the gas temperature and $C$ is equal to a dimensional numerical
constant which depends on the single particle mass.

In this work we have done the first set of measurements to prove Eq. (6) in
a situation where $\omega _{0}$ is kept constant. This is equivalent to an
isocoric transformation of the trapped gas. As the evaporation takes place
we record $\sigma _{\text{x}}$ and $\sigma _{\text{z}}$ and estimate $\sigma
_{\text{y}}$, by the relation between the oscillation frequencies. Thus with
the total number of particle and gas temperature we create a plot of $\frac{P%
}{\text{N}C}$ as a function of T, as shown in figure 8, previously to reach
BEC.

\FRAME{ftbpFU}{4.0179in}{2.9905in}{0pt}{\Qcb{Graphic showing the evolution
of normalized harmonic pressure $\frac{\text{P}}{\text{NC}}$ as a function
of temperature T.}}{}{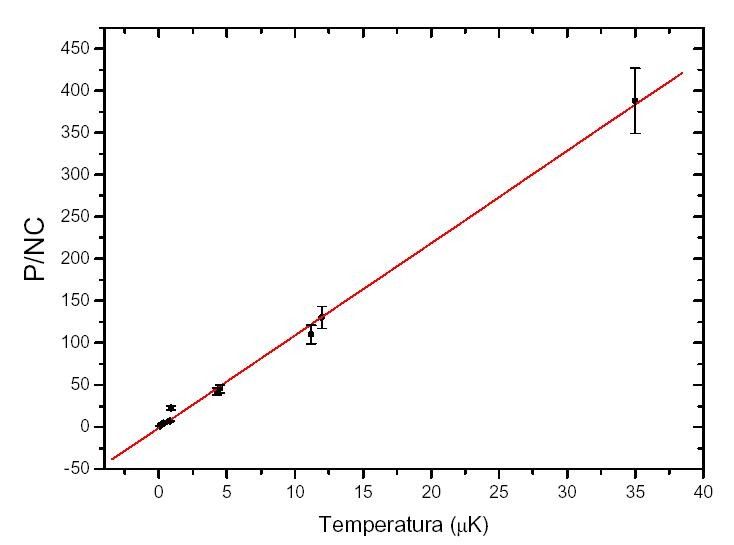}{\special{language "Scientific Word";type
"GRAPHIC";maintain-aspect-ratio TRUE;display "USEDEF";valid_file "F";width
4.0179in;height 2.9905in;depth 0pt;original-width 7.6458in;original-height
5.6775in;cropleft "0";croptop "1";cropright "1";cropbottom "0";filename
'figure8.jpg';file-properties "XNPEU";}}

The observed linearity of $\frac{P}{\text{N}C}$ with T shows that $PV$ $%
\propto $ T, as expected since the limits presented by V. Romero-Rochin\cite%
{vitor} in his work is the classical limit where the defined pressure and
volume would obey the ideal gas state equation. We would like to explore the
deviation of the state equation from the ideal gas case. That will allow us
to finally evaluate the heat capacity and other important thermodynamic
functions as the gas passes from the classical to the quantum regime. This
work is now under preparation. For this purpose is extremely important the
detection of the atomic distribution as the gas is still trapped (\textit{in
situ} detection). A full investigation of harmonic pressure under isothermal
or isochoric transformation will be implemented in our laboratory.

\section{Conclusions}

In this work we demonstrate that we have achieved Bose-Einstein condensation
in a sample of Na atoms by \textit{in situ} observation of the atomic
density profile and temperature, resulting in the evaluation of the
phase-space density. The overall analysis allows us to understand that the
final low number of atoms was consequence on the fact that runaway
evaporation only took place at the final stage of the process. The route to
BEC was also analyzed on a thermodynamic view showing the evolution of the
gas pressure as the temperature was lowered. The \textit{\ in situ}
observation of the gas will allow us further thermodynamical analysis of the
system as it approaches BEC.

We would like to thank to Daniel V. Magalh\~{a}es, Phillipe W. Courteille,
Robin Kaiser and Kristian Helmerson by helpful discussions and technical
support. We appreciate financial support from Fapesp, CNPq and CAPES
Brazilian agencies and MM\ Optics.

\bigskip 

\end{document}